\documentclass[12pt]{article}
\usepackage{graphicx}

\newcommand\pubnumber{DPF2015-315}
\newcommand\pubdate{\today}

\def\unl{Department of Physics and Astronomy\\
University of Nebraska-Lincoln, Lincoln, NE 68588-0299}

\def\Title#1{\begin{center} {\Large #1 } \end{center}}
\def\Author#1{\begin{center}{ \sc #1} \end{center}}
\def\Address#1{\begin{center}{ \it #1} \end{center}}

\newcommand\pubblock{\rightline{\begin{tabular}{l} \pubnumber\\
         \pubdate  \end{tabular}}}
\newenvironment{Abstract}{\begin{quotation}  }{\end{quotation}}
\newenvironment{Presented}{\begin{quotation} \begin{center} 
             PRESENTED AT\end{center}\bigskip 
      \begin{center}\begin{large}}{\end{large}\end{center} \end{quotation}}

\def\beq{\begin{equation}}
\def\eeq#1{\label{#1}\end{equation}}
\def\eeqn{\end{equation}}

\def\beqa{\begin{eqnarray}}
\def\eeqa#1{\label{#1}\end{eqnarray}}
\def\eeqan{\end{eqnarray}}

\let\bar=\overbar

\def\Dslash{\not{\hbox{\kern-4pt $D$}}}
\def\dslash{\not{\hbox{\kern-2pt $\del$}}}

\def\msb{{\bar{\ssstyle M \kern -1pt S}}}

\begin{document}
\begin{titlepage}
\pubblock

\vfill
\Title{Run-1 Single-top measurements at CMS}
\vfill
\Author{Rebeca Gonzalez Suarez}
\Address{\unl}
\vfill
\begin{Abstract}
The Run-1 of the LHC was very successful for single top physics. The main single top production mode, $t-$channel, is now well established. The $t-$channel cross-section was measured with unprecedented precision and $t-$channel events have been used for the first time to perform measurements of standard model (SM) properties, such as: $|V_{tb}|$, top quark polarization, or $W$-helicity fractions. The associated production with a $W$ boson, $tW$, has a large cross section at the LHC and has been observed by CMS for the first time. Finally, the $s-$channel also has been studied and limits set on its production cross section. Single top results produced by the CMS experiment at 7 and 8~TeV are presented in the following.
\end{Abstract}
\vfill
\begin{Presented}
DPF 2015\\
The Meeting of the American Physical Society\\
Division of Particles and Fields\\
Ann Arbor, Michigan, August 4--8, 2015\\
\end{Presented}
\vfill
\end{titlepage}
\def\thefootnote{\fnsymbol{footnote}}
\setcounter{footnote}{0}

\section{Introduction}

At the LHC, top quarks are produced mainly in $t\bar{t}$ pairs via strong interaction. Single top quark production is an alternative mode that happens at a lower rate via electroweak interaction. There are three main single top process: $t-$channel, $tW$ associated production, and $s-$channel.

The Tevatron experiments observed top quarks via $t\bar{t}$ production for the first time in 1995, and in 2009 via single top; but the LHC has been competitive since starting running, publishing the first single top paper in 2011.

The Run-1 of the LHC lasted three years and delivered about 5$fb^{-1}$ of proton proton collisions at 7~TeV and about 20$fb^{-1}$ at 8TeV. During that time, the CMS experiment registered more than 5 million of $t\bar{t}$ pairs, around 2 million of single top quarks via $t-$channel, half a million of $tW$ events, and a bit more than 100 thousand $s-$channel events.

\section{Main single top production modes, inclusive cross sections}

\subsection{$t-$channel}

The single top $t-$channel production mode is the process with the highest cross section at both the Tevatron and the LHC. The final state studied in this channel is a lepton+jets signature. The signal characterized by one isolated muon or electron; missing transverse energy ($\not\!\!{E_{T}}$); a central jet coming from the decay of a b-hadron; a light-quark jet from the hard scattering process, that is often forward; and additionally,  a second b jet produced in association to the top quark. The main backgrounds for the analysis are $W$+jets, $t\bar{t}$ and multijet events. 
\\
\\
At 7~TeV CMS measured the inclusive $t-$channel cross section using multivariate methods (BDT, NN) and the shape of the pseudorapidity of the light jet, $|\eta_{j'}|$~\cite{Chatrchyan:2012ep}. The analysis was performed in different regions defined by the number of jets, using b-tagging. Multijet and $W$+jets backgrounds were estimated using data. At 7~TeV, the statistical, systematic, and theory uncertainties where on the same level. 

At 8~TeV, the measurement was done using the $|\eta_{j'}|$ analysis alone~\cite{Khachatryan:2014iya}. A similar approach as for 7~TeV was used, and $t\bar{t}$ background was also estimated using a data-driven method, as in the case of $W$+jets and multijet. The full integrated luminosity at 8~TeV was used.  The analysis is not limited statistics, and the the main systematic uncertainties come from signal modelling (6\%) and jet energy scale/resolution and $\not\!\!{E_{T}}$(4\%).

The measured values of the $t-$channel inclusive cross section at 7 and 8~TeV are presented in Table~\ref{tab:tchan} together with the SM predictions at NLO. 

An LHC combination was made at 8TeV using the results by ATLAS and CMS. This first single top combination of the LHC, had as a result a cross section value of  $\sigma_{t-channel}^{8~TeV}$=85$\pm$4(stat)$\pm$11(syst)$\pm$3(lumi)~pb, in good agreement with the SM prediction. 
\begin{table}[htb]
\begin{center}
\begin{tabular}{l|cc}  
 $\sqrt{s}$ &  Measured [pb]  &  Predicted [pb] \\ \hline
7~TeV &  67.2$\pm$3.7(stat)$\pm$3.0(syst)$\pm$3.5(th)$\pm$1.5(lumi) &  63.89+2.91-2.52 \\
8~TeV & 83.6$\pm$2.3(stat)$\pm$7.4(syst) & 84.69+3.76-3.23 \\
\hline
\end{tabular}
\caption{Measured and predicted values of the inclusive production cross section of single top $t-$channel. The predicted values correspond to NLO calculations.}
\label{tab:tchan}
\end{center}
\end{table}

\subsection{tW associated production}

The associated production of a single top quark with a $W$ boson, $tW$, that was not accessible at the Tevatron due to its small cross section, has the second largest cross section at the LHC.

The final state studied is a dilepton signature. Signal events are characterized by two opposite-sign, isolated leptons, a substantial $\not\!\!{E_{T}}$ due to having 2 neutrinos in the final state, and a jet coming from a b decay. The main backgrounds are $t\bar{t}$, which represents the main analysis challenge, followed by Drell-Yan processes in the same-flavor final states.

This process could be studied for the first time at the LHC, but it is still a difficult process due to its cross section and background contamination. With 7~TeV data, CMS reported evidence for the process~\cite{Chatrchyan:2012zca}. With the full integrated luminosity recorded by CMS at  7~TeV, 4.9$fb^{-1}$, the observed $tW$ signal had a significance of 4.0~$\sigma$.

At 8~TeV the process was finally observed by CMS with a significance higher than 5$\sigma$~\cite{Chatrchyan:2014tua}. At 7~TeV a multivariate discriminant (BDT) was used to separate $tW$ signal from $t\bar{t}$ background, a cross check analysis that used kinematic requirements was also performed. At 8~TeV, a more sophisticated BDT was designed, and, in addition to the cross check made with kinematic requirements, a fit on a discriminant variable was also performed as intermediate step. As a result, the 8~TeV analysis was already not statistically limited, and with an integrated luminosity of 12.2$fb^{-1}$ the analysis already observed the $tW$ signal with a significance of 6.1$\sigma$.

The main challenge for this final state is the contamination from $t\bar{t}$ events. When tt events have one jet outside the acceptance or misreconstructed, they mimic perfectly the $tW$ signal. And not only they have very similar final states, the $tW$ and $t\bar{t}$ diagrams mix at NLO. Therefore, this background is the most important for the study of $tW$ and the main uncertainties come from its modeling. In both analysis at 7 and 8 TeV, this problem was approached by introducing two independent $t\bar{t}$ control regions, defined by the number of jets and b-tags. In the case of the 8 TeV analysis, the use of variables related to `loose' jets, defined as jets passing a set of quality criteria slightly looser than the main jets used in the analysis, allowed for a better control of this background.

The measured values of the single top $tW$ associated production inclusive cross section at 7 and 8 TeV are presented in Table~\ref{tab:tw} together with the SM predictions.

\begin{table}[htb]
\begin{center}
\begin{tabular}{l|cc}  
 $\sqrt{s}$ &  Measured [pb]  &  Predicted [pb] \\ \hline
7~TeV &  16+5-4 & 15.6$\pm$0.4$\pm$1.1 \\
8~TeV & 23.4$\pm$5.4 & 22.2$\pm$0.6$\pm$1.4 \\
\hline
\end{tabular}
\caption{Measured and predicted values of the inclusive cross section of single top $tW$ associated production.}
\label{tab:tw}
\end{center}
\end{table}

An ATLAS and CMS tW combination was performed at 8TeV with a result of $\sigma_{tW}^{8~TeV} = 25.0\pm1.4(stat)\pm4.4(syst)\pm0.7(lumi)$pb.

\section{$s-$channel}

The $s-$channel is the mode with the lowest cross section at the LHC. It was more important at the Tevatron, where the study of data after the shutdown allowed for the observation of the process~\cite{CDF:2014uma}.

The $s-$channel it is the most sensitive single top mode to new physics, sharing topology with $W'$ bosons, or charged Higgs bosons. However, it has a very challenging final state, with a low production cross section, and very difficult to separate from background.

The signal signature studied is lepton+jets, with a lepton, electron or muon, and $\not\!\!{E_{T}}$ from the decay of a $W$ boson, and two jets with high transverse momentum originating from b-quarks. The main background comes from $t\bar{t}$, and $W+$jets and multijet, amongst other, also contribute.

At 8~TeV, CMS has a preliminary result~\cite{CMS:2013fmk}. The analysis uses the full luminosity and final states with electrons and muons, and it is based on a
BDT to separate $t\bar{t}$ from $s-$channel signal. However, the sensitivity is still very limited, only an upper limit of 2.1 times the SM is set on $s-$channel production. The
estimated value of the cross section is $\sigma_{s-channel}^{8~TeV} = 6.2\pm5.4(exp)\pm5.9(th) =6.2+8.0-5.1$pb, to be compared with the SM prediction of 5.55$\pm$0.08(scale)$\pm$0.21(pdf)~pb at NNLL.

With the running conditions of the Run-2 of the LHC, the sensitivity towards $s-$channel will be difficult to improve. The best signal-over-background ratio that can be achieved for this channel in terms of production cross section at the LHC corresponds to a center of mass energy of 7~TeV.

\section{Measurement of properties in single top signatures}

At the LHC, the single top production is large enough to allow for the measurement of top properties in single top signatures. This is complementary to the measurements performed using $t\bar{t}$ events, provides an additional handle to test potential BSM phenomena, and it is valuable to get the full picture of the top quark.

\subsection{Cross section and $|V_{tb}|$}

\begin{figure}[htb]
\centering
\includegraphics[height=3in]{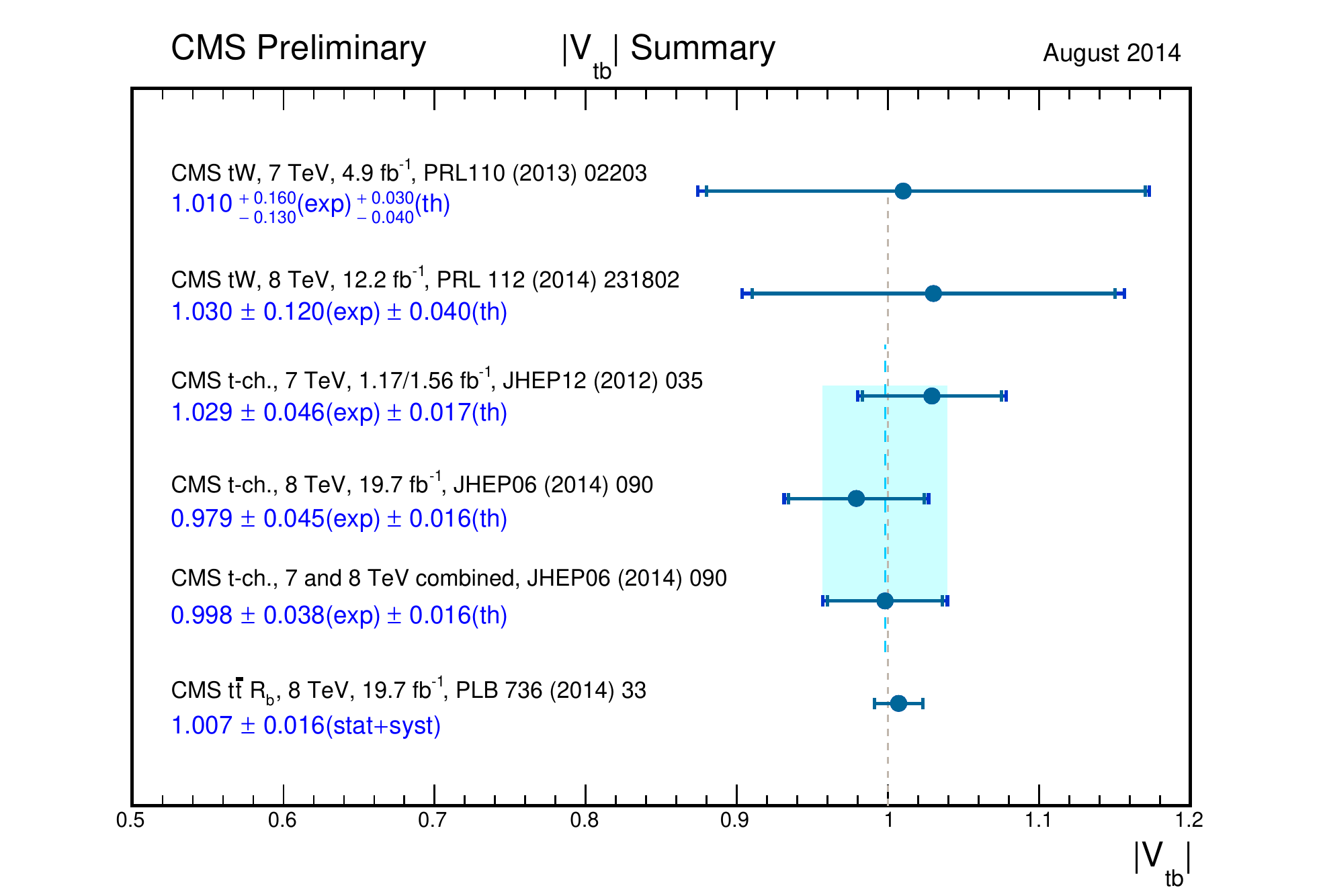}
\caption{Summary of $|V_{tb}|$ values extracted from single top and $t\bar{t}$ measurements in CMS.}
\label{fig:vtb}
\end{figure}

From the inclusive production cross section of single top in both $t-$channel and $tW$, a value of the CKM matrix element, $|V_{tb}|$, can be extracted. Considering $|V_{td}|$ and $|V_{ts}|$ to be much smaller than $|V_{tb}|$, the cross section of the single top production relates to $|V_{tb}|^{2}$. The summary of the $|V_{tb}|$  values extracted from the different single top process at the LHC, including CMS is shown in Figure~\ref{fig:vtb}. All the values are in agreement with the SM expectation of $|V_{tb}|=1$.

\subsection{top/anti-top asymmetry, R}

Within the measurement of the $t-$channel cross-section at 8 TeV~\cite{Khachatryan:2014iya}, the top/anti-top asymmetry, R, can be measured. Due to the relative proportion of $u$ and $d$ quarks in the proton, more tops than anti-tops are expected to be produced. Therefore, the measurement of R can give valuable information on the internal structure of the proton.

\begin{table}[htb]
\begin{center}
\begin{tabular}{l|cc}  
 &  Measured [pb]  &  Predicted [pb] \\ \hline
$\sigma_{t}$ &  53.8$\pm$1.5(stat)$\pm$4.4(sys) & 54.87+2.29-1.94 \\
$\sigma_{\bar{t}}$ & 27.6$\pm$1.3(stat)$\pm$4.4(sys) & 29.74+1.67-1.51 \\
\hline
\end{tabular}
\caption{Measured and predicted (NLO) values of the top and anti-top quark production via $t-$channel at 8~TeV.}
\label{tab:R}
\end{center}
\end{table}

The measured values of $\sigma_{t}$ and $\sigma_{\bar{t}}$  are presented in Table~\ref{tab:R}, the value of the top/anti-top asymmetry that is obtained from them is $R=1.95\pm0.10~(stat)\pm0.19~(sys)$.

\subsection{Differential measurements}

At 8~TeV, CMS measured the $t-$channel differential cross section~\cite{CMS:2014ika} using the full luminosity and a general selection similar to the inclusive cross section selection~\cite{Khachatryan:2014iya}. The analysis then uses a NN to isolate a purer t-channel sample, selecting events with a NN discriminant value above an optimal value. The distributions of the $p_{T}$ and $\eta$ of the top quarks are corrected for detector effects and compared directly with different theoretical predictions: POWHEG+Pythia, aMC@NLO+Pythia, and CompHEP. All the predictions agree with the data within uncertainties.

A measurement of the $W-$helicity fractions was also performed using $t-$channel events at 8~TeV~\cite{Khachatryan:2014vma}. This measurement uses the $\cos{\theta^{\star}}$ distribution, after applying the exact same selection and background estimation as the standard $t-$channel inclusive cross section measurement. The angle $\theta^{\star}$ is defined as the angle between the $W$ boson in the top rest frame and the lepton in the $W$ rest frame, and relates directly to the $W-$helicity fractions. The values obtained longitudinal, $F_{0}$, left-handed, $F_{L}$, and right-handed, $F_{R}$, helicity fractions, are presented in Table~\ref{tab:Whel}.

\begin{table}[htb]
\begin{center}
\begin{tabular}{l|cc}  
 &  Measured  &  Predicted  \\ \hline
$F_{L}$ &  0.298$\pm$0.028(stat)$\pm$0.032(sys)  & 0.311$\pm$0.005  \\
$F_{R}$ & -0.018$\pm$0.019(stat)$\pm$0.011(sys)  & 0.0017$\pm$0.0001  \\
$F_{0}$ & 0.720$\pm$0.039(stat)$\pm$0.037(sys)  & 0.687$\pm$0.005 \\
\hline
\end{tabular}
\caption{Measured and predicted (NNLO) values of the W helicity fractions.}
\label{tab:Whel}
\end{center}
\end{table}

Single top quarks are highly polarized with the spin aligned with the recoiling light jet. The top quark polarization relates to the spin asymmetry as $A_{l} =\frac{1}{2}P_{t}\alpha_{l}$, and $A_{l}$ can be extracted from the $\cos{\theta^{\star}_{l}}$ distribution. A measurement of the top quark polarization is made at 8 TeV~\cite{CMS:2013rfa} using a BDT to get a $t-$channel enriched sample. The asymmetry is obtained from the unfolded distributions in the $e$ and $\mu$ channels separately and combined using BLUE. The measured value of the asymmetry is $A_{l} = 0.41\pm0.06(stat)\pm0.16(sys)$, that compares well with the SM expectation of 0.44, and from it, a value of $P_{t} = 0.82\pm0.12(stat)\pm0.32(sys)$ is obtained.

\section{Search for FCNC and Anomalous Couplings}

Limits on anomalous $Wtb$ couplings can be extracted from SM measurements like the $W-$helicity fractions, or the top polarization. Additionally, dedicated analyses searching for deviations from the SM are also in place in single top signatures.

At 7~TeV, a search for FCNC $tZ$ exploring a three-lepton signature was performed by CMS~\cite{CMS:2013nea}. For this, simulated samples with four different anomalous scenarios were produced. For each of the scenarios, $gut$, $gct$, $Zut$, and $Zct$, a BDT was trained against SM $tZq$. No excess was observed with respect to the SM expectations and limits were set on couplings and branching fractions.

The $tq\gamma$ production, single top produced in association with a photon, was also explored. In this case a potential enhancement on the branching ratio of $t\rightarrow u(c)\gamma$ due to FCNC was studied. The analysis was done at 8~TeV using final states with $\mu$ only~\cite{CMS:2014hwa}. Samples with anomalous $tu\gamma$ and $tc\gamma$ couplings were produced and dedicated BDTs were trained for each scenario. No excess was observed and limits on couplings and branching fractions were set.

Finally, a dedicated search for FCNC and anomalous couplings in $t-$channel was also performed~\cite{CMS:2014ffa} with 7~TeV data and in $\mu$ final states only. Several anomalous operators in the $Wtb$ vertex and $tcg$ and $tug$ FCNC couplings were explored and dedicated samples produced. The analysis then used neural networks to separate different scenarios considered from the SM. As in the previous, no excess was found above the SM expectations and limits on couplings and branching fractions were established.

\section{Summary}

Single top signatures, largely unknown until recently, have been largely explored by the LHC experiments during the Run-1. In particular, CMS has studied the three main production modes, establishing the $t-$channel and $tW$ associated production, and exploring the complicated $s-$channel processes. Single top quarks produced via $t-$channel have been already used for measurements, measuring the $W-$helicity fractions, top polarization, or the CKM matrix element $|V_{tb}|$. Conventional and rare single top production modes have been explored in the search for BSM physics: FCNC and Anomalous Couplings.

Run-2 will be the time to fully explore single top signatures, in particular to look for physics beyond the standard model.

\end{document}